\documentclass[final]{IEEEtran}

\newlength \figwidth
\usepackage{cite}
\ifCLASSINFOpdf
  \usepackage[pdftex]{graphicx}
	\usepackage{epstopdf}
  \graphicspath{{../Figures/}}
  \DeclareGraphicsExtensions{.eps}
\else
  \usepackage[dvips]{graphicx}
\fi
\usepackage[cmex10]{amsmath}
\usepackage{amssymb}
\usepackage{amsthm}
\usepackage{url}
\usepackage{tabulary}
\usepackage{multirow}
\usepackage{color}

\usepackage{color}
\usepackage{times}
\usepackage{verbatim}
\usepackage{floatflt}
\usepackage{enumerate}
\usepackage{array}
\usepackage{multicol,afterpage,wrapfig} 
\usepackage{tikz}
\usepackage{algorithm}
\usepackage[noend]{algpseudocode}
\makeatletter
\def\BState{\State\hskip-\ALG@thistlm}
\makeatother
\usepackage{amsmath,amssymb,eucal}
\usepackage[utf8]{inputenc}
\usepackage{epsfig}
\usepackage{exscale}
\usepackage{latexsym}
\usepackage{verbatim}
\usepackage{amsfonts}
\usepackage{helvet}
\usepackage{multirow}
\usepackage{cite}
\usepackage{balance}
\usepackage{color}
\usepackage{transparent}
\usepackage{import}
\usepackage{mathtools}
\usepackage{tikz}
\usetikzlibrary{automata,arrows,positioning,calc}
\usepackage{bbm}
\usepackage[printonlyused,withpage]{acronym}
\usepackage{tabu}
\usepackage{longtable}
\usepackage{balance}
\usepackage{footnote}
\usepackage{tablefootnote}
\usepackage[font=footnotesize]{caption}
\usepackage{enumitem}

\def\BibTeX{{\rm B\kern-.05em{\sc i\kern-.025em b}\kern-.08em
    T\kern-.1667em\lower.7ex\hbox{E}\kern-.125emX}}
    
\renewcommand{\IEEEQED}{\IEEEQEDopen}

\newcommand*\xbar[1]{%
  \hbox{%
    \vbox{%
      \hrule height 0.5pt % The actual bar
      \kern0.36ex%         % Distance between bar and symbol
      \hbox{%
        \kern-0.12em%      % Shortening on the left side
        \ensuremath{#1}%
        \kern-0.12em%      % Shortening on the right side
      }%
    }%
  }%
}

\newfont{\bbb}{msbm10 scaled 500}

\newfont{\bb}{msbm10 scaled 1100}

\newcommand{\executeiffilenewer}[3]{%
\ifnum\pdfstrcmp{\pdffilemoddate{#1}}%
{\pdffilemoddate{#2}}>0%
{\immediate\write18{#3}}\fi%
}

\usetikzlibrary{arrows,calc}
\allowdisplaybreaks

\IEEEoverridecommandlockouts
\begin{document}
\pagenumbering{gobble}
%\makeatletter
%\def\thm@space@setup{\thm@preskip=0pt
%\thm@postskip=0pt}
%\makeatother

\newtheorem{Theorem}{\bf Theorem}
\newtheorem{Corollary}{\bf Corollary}
\newtheorem{Remark}{\bf Remark}
\newtheorem{Lemma}{\bf Lemma}
\newtheorem{Proposition}{\bf Proposition}
\newtheorem{Assumption}{\bf Assumption}
\newtheorem{Definition}{\bf Definition}
%\title{Trade-off Analysis Among Users Multiplexing, Channel Training Interference and Overhead in UL SRS Design for NR Massive MIMO Systems}
%\title{UL SRS Coordination in \\ 5G Massive MIMO Systems}
\title{Uplink Sounding Reference Signal Coordination to Combat Pilot Contamination in 5G Massive MIMO}
\author{\IEEEauthorblockN{{Lorenzo~Galati~Giordano$^{\star}$, Luca~Campanalonga$^{\dag \star}$, David~L\'{o}pez-P\'{e}rez$^{\star}$, Adrian~Garcia-Rodriguez$^{\star}$, Giovanni~Geraci$^{\star}$, Paolo~Baracca$^{\diamond}$, and Maurizio~Magarini$^{\dag}$}}\\
\normalsize\IEEEauthorblockA{\emph{$^{\star}$Nokia Bell Labs Ireland, $^{\diamond}$Nokia Bell Labs Germany, $^{\dag}$Politecnico di Milano Italy}}}
\maketitle
\thispagestyle{empty}

\IEEEpeerreviewmaketitle
\begin{abstract}

To guarantee the success of massive multiple-input multiple-output (MIMO), one of the main challenges to solve is the efficient management of pilot contamination.
Allocation of fully orthogonal pilot sequences across the network would provide a solution to the problem, but the associated overhead would make this approach infeasible in practical systems. Ongoing fifth-generation (5G) standardisation activities are debating the amount of resources to be dedicated to the transmission of pilot sequences, focussing on uplink sounding reference signals (UL SRSs) design. In this paper, we extensively evaluate the performance of various UL SRS allocation strategies in practical deployments, shedding light on their strengths and weaknesses. Furthermore, we introduce a novel UL SRS fractional reuse (FR) scheme, denoted neighbour-aware FR (FR-NA). The proposed FR-NA generalizes the fixed reuse paradigm, and entails a trade-off between $i)$ aggressively sharing some UL SRS resources, and $ii)$ protecting other UL SRS resources with the aim of relieving neighbouring BSs from pilot contamination. Said features result in a cell throughput improvement over both fixed reuse and state-of-the-art FR based on a cell-centric perspective.

\end{abstract}
\section{Introduction}

In order to achieve a high spectral efficiency, massive multiple-input multiple-output (MIMO) has been identified as one of the most promising technologies for the fifth-generation (5G) wireless communication systems \cite{Tom2014Magazine, Tom2015BLTJ,GarGerGal2017}. Massive MIMO base stations (BSs) employ a large number of radiating antenna elements to spatially multiplex a significant number of users (UEs) while providing large beamforming gains. One of the most important operations in massive MIMO is channel training, i.e., the acquisition of precise channel state information (CSI). Such acquisition process relies on the use of reference signals. In a time division duplex (TDD) system, downlink (DL) channels can be estimated through uplink (UL) pilots thanks to channel reciprocity. This makes the training time proportional to the number of UEs, rendering TDD particularly suitable for massive MIMO \cite{1608543_TDDFDDMarzetta}.

\subsection{Background and Motivation} %UL SRS

In the fourth-generation (4G) wireless communication systems, also known as Long Term Evolution (LTE), UL CSI is acquired through the use of sounding reference signals (SRSs). 
Various SRS configurations are possible, resulting in different channel estimation capabilities in both the time and frequency domains \cite{3GPP36211}. Although SRSs are mostly used in LTE for providing wide- or sub-band CSI -- thus assisting the UL medium access control (MAC) scheduler in allocating UEs to resource blocks (RBs) --, SRSs are considered in the ongoing 5G standardisation as the main candidate to carry UL massive MIMO pilots in TDD systems. In this context, the assignment of UEs to UL SRSs has become critical, as it can significantly affect the system performance. Indeed, since the number of UL SRSs for a given bandwidth is limited, and because the number of multiplexed UEs may be significantly larger than that in existing systems, UL SRSs must be reused across cells. Said necessary UL SRS reuse results in pilot contamination, which has been identified as one of the main limiting factors in massive MIMO systems \cite{Tom2014Magazine,Tom2015BLTJ}.

Several works have been carried out in the past few years to address the issue of pilot contamination, and a comprehensive survey can be found in \cite{7339665_SurveyPilotCont}. 
In particular, an efficient approach to mitigate pilot contamination consists in coordinating the use of the UL SRSs across network cells by applying fixed pilot reuse schemes \cite{7185106_ReuseSchemeMarzetta}. Through these schemes, UL SRSs are assigned such that they are not reused by any neighbouring cell in a given area. Meeting such requirement comes at the expense of increasing the associated training overhead, as more time-frequency resources are needed to train the same number of UEs. Fractional reuse (FR) has been proposed as a generalisation of the fixed pilot reuse approach \cite{7247312_FractionalPilotReuseDebbah,7127059_Hanzo_SoftPilotReuse}. 
In FR, some UEs benefit from a coordinated pilot assignment and are relieved from severe pilot contamination, while the rest of the UEs aggressively reuse the remaining pilots. For  example, in \cite{7247312_FractionalPilotReuseDebbah,7127059_Hanzo_SoftPilotReuse}, vulnerable cell-edge UEs are assigned a subset of protected pilots, while cell-centre UEs share the same non-protected pilots across the network. While this approach has the merit of refining the trade-off between pilot contamination and training overhead, we show that the metric it employs -- based on a cell-centric approach -- leaves room for improvement.

\subsection{Approach and Summary of Results} %This paper

In this paper, we embrace the FR paradigm and propose a novel neighbour-aware FR (FR-NA) approach for UL SRS coordination. Unlike cell-centric FR (FR-CC) in \cite{7247312_FractionalPilotReuseDebbah}, in the proposed FR-NA BSs follow a selfless approach, driven by the pilot contamination each of their associated UEs causes at neighbouring cells. We also provide an in-depth performance evaluation for FR-NA, comparing it to conventional fixed reuse schemes and to FR-CC for a variety of SRS configurations and under realistic channel assumptions. Our study contributes the following insights:
\begin{itemize}
	\item Randomly assigning UL SRS sequences without BS coordination results in poor performance regardless of the number of available sequences. This occurs either because pilot contamination is not prevented or because the associated overhead is large.
	\item Selection of the appropriate UL SRS reuse scheme is key to guarantee a satisfactory performance, and it depends on the number of active UEs and on the available time-frequency resources for UL SRS transmission. 
	\item The proposed FR-NA UL SRS allocation scheme provides $43\%$ DL BS throughput gains with respect to Reuse~1, i.e., fixed full reuse, where all UL SRS resources are reused by all BSs. 
\end{itemize}

%\clearpage
%\vspace*{-0.3cm}
\section{System Model} \label{MMIMOSystemModel}

We consider a cellular system with a set $\mathcal{B}$ of $N_\mathrm{B}$ BSs deployed on a hexagonal layout, where each site is formed by 3 different co-located BSs, each forming a sector.
Each BS $b$ is equipped with a massive MIMO array with $N_\mathrm{A}$ antenna elements and we consider that $N_\mathrm{U}$ single-antenna UEs are uniformly distributed in the geographical area. The set of scheduled UEs in each BS $b$ per transmission time interval (TTI) and its cardinality are denoted by $\mathcal{K}_b$ and $\vert \mathcal{K}_b \vert = N_{\mathrm{K},b}, \forall b \in \left\{ 1, \ldots, N_\mathrm{B} \right\}$, respectively. The system is operated in a TDD fashion, where channels are estimated at the BS through the use of UL SRSs sent by the UEs under the assumption of channel reciprocity \cite{Tom2015BLTJ}.

\subsection{Subframe Structure}

\begin{figure}[t]
	\includegraphics[width=\columnwidth]{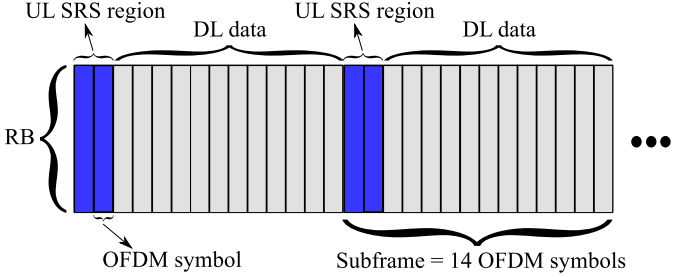}
	\caption{Subframe structure model. Dark blue zones are dedicated to UL SRS transmission for all scheduled UEs, whereas light grey zones are dedicated to DL data transmission.}
	\centering
	\label{fig:FrameStructure}
\end{figure}

In this paper, we focus on how UL SRS coordination affects the DL data rate. Fig. \ref{fig:FrameStructure} defines the subframe structure adopted in this work, 
which is formed by $T=14$ OFDM symbols and a certain number of RBs, 
according to the adopted bandwidth (e.g., 100 RBs for the 20 MHz case) \cite{3GPP36211}. 
Within this subframe structure, 
a variable number $\tau$ of OFDM symbols (dark blue area in Fig. \ref{fig:FrameStructure}) is dedicated to the transmission of UL SRSs, 
while the remaining $T-\tau$ ones (light grey area in Fig. \ref{fig:FrameStructure}) are allocated to DL data.
It is important to note that in our system model 
\emph{i)} the value of $\tau$ is kept constant for all BSs in the network, 
but can change according to the different UL SRS reuse schemes,
and that \emph{ii)} the UEs trained in the UL SRS region of size $\tau$ OFDM symbols are those which are served immediately after in the DL data region of size $T-\tau$ OFDM symbols.
We note that this versatile subframe structure can be easily adapted to the 5G frame structure by simply setting the appropriate subframe numerology and partitioning. 

\subsection{UL SRS and Channel Estimation}

We adopt the UL SRS framework of LTE, where various UL SRSs configurations are possible as specified in \cite{3GPP36211}. We focus on UL SRSs that span the entire bandwidth to allow a wideband channel sounding. With this in mind, and considering the 2$\times$ time repetition factor as well as the 8 possible cyclic shifts of the UL SRSs, a maximum of 16 UL SRSs can be multiplexed in one OFDM symbol \cite{3GPP36211}. 5G standardisation is currently discussing whether amendments to this framework are needed to effectively support massive MIMO applications and an increasing number of UEs multiplexed per TTI, e.g., by constraining the number of OFDM symbols dedicated to UL SRSs.

Let $\pmb{\phi}_{k,b} \in \mathbb{C}^{N_{\mathrm{P}} \times 1}$ be the pilot sequence of the $k$-th UE in the $b$-th cell. Here, $N_{\mathrm{P}}$ denotes the pilot sequence length, which corresponds to the total number of UL SRSs available.
Let us also define $\mathbf{\Phi}_b$ = $[\pmb{\phi}_{1,b}, \pmb{\phi}_{2,b}, ..., \pmb{\phi}_{N_{\mathrm{K},b},b}]^{\mathrm{T}} \in \mathbb{C}^{N_{{\mathrm{K}},b} \times N_{\mathrm{P}}}$
as the composite matrix containing the UL SRS sequences used by the $N_{{\mathrm{K}},b}$ UEs scheduled by BS $b$. Without loss of generality, in our system model \emph{i)} one UE $k$ is allocated at most one UL SRS sequence, and \emph{ii)} the maximum number $N_{\mathrm{K},b}$ of UEs scheduled per TTI cannot be larger than $N_{\mathrm{P}}$.
Moreover, since UL SRSs used within the same BS cell are orthogonal and cannot be reused,
it must be $\mathbf{\Phi}_b \mathbf{\Phi}_b^{\mathrm{H}} = \mathbf{I}_{N_{\mathrm{K},b}}$. The UL SRSs signals $\mathbf{Y}_b \in \mathbb{C}^{N_{\mathrm{A}} \times N_{\mathrm{P}}}$ received at BS $b$ per time-frequency resource over the $N_{\mathrm{A}}$ antennas can thus be expressed as \cite{7127059_Hanzo_SoftPilotReuse}
\begin{equation}
	\mathbf{Y}_b = \sqrt{\rho} \sum_{j=1}^{N_\mathrm{B}} \mathbf{H}_{b,j} \mathbf{\Phi}_j + \mathbf{N}_b,
	\label{eq:pilot}
\end{equation}
where $\rho$ is the UL transmit power applied by a UE to its UL SRS, which we assume identical for all UEs, 
$\mathbf{H}_{b,j} \in \mathbb{C}^{N_{\mathrm{A}} \times N_{\mathrm{K},b}}$ is the UL channel matrix between BS $b$ and the UEs scheduled by BS $j$, 
and $\mathbf{N}_b \in \mathbb{C}^{N_{\mathrm{A}} \times N_{\mathrm{P}}}$ is the UL channel additive white Gaussian noise (AWGN) matrix formed by independent and identically distributed (i.i.d.) Gaussian entries with variance $\sigma^{2}$. 

%\subsection{Channel Estimation}

After receiving the UL SRSs, 
each BS $b$ estimates the UL channel matrix from the received UL SRSs, 
which include the contamination effects caused by the UEs scheduled by all neighbouring BS $j$ reusing the same UL SRSs.
Considering least-squares estimation, the resultant uplink channel estimate $\widehat{\mathbf{H}}_{b} = [\widehat{{\mathbf{h}}}_{1,b},\ldots,\widehat{\mathbf{h}}_{N_{\mathrm{K},b},b}] \in \mathbb{C}^{N_{A} \times N_{\mathrm{K},b}}$ of the $b$-th BS can be written as \cite{6493983_MultiMIMOULNgoMarzetta}
\begin{equation}
	\widehat{\mathbf{H}}_{b} = \dfrac{1}{\sqrt{\rho}} \mathbf{Y}_b \mathbf{\Phi}_b^{\mathrm{H}} = \mathbf{H}_{b,b} + \sum^{N_\mathrm{B}}_{j \in \mathcal{B} \backslash b} \mathbf{H}_{b,j} + \dfrac{1}{\sqrt{\rho}}\mathbf{N}_b \mathbf{\Phi}_b^{\mathrm{H}}.
	\label{eq:channel_estimation}
\end{equation} 
%[David]: I am confused about \mathbf{H}}_{b,b}  and  \mathbf{H}}_{b}. Are they different or the same?

\subsection{Zero-Forcing Precoding and Throughput Formulation}

From the channel estimate in \eqref{eq:channel_estimation}, the zero forcing (ZF) precoder
\begin{equation}
	\mathbf{W}_b^{\mathrm{ZF}} = \left[ {\mathbf{w}}_{1,b}^{\mathrm{ZF}},\ldots,{\mathbf{w}}_{N_{\mathrm{K},b}, b}^{\mathrm{ZF}} \right]	
\end{equation}
at BS $b$ can be calculated as \cite{1261332_ZF}
\begin{equation}
	\mathbf{W}_b^{\mathrm{ZF}} = \left(\mathbf{D}_b^{\mathrm{ZF}}\right)^{-\frac{1}{2}} \widehat{\mathbf{H}}_b \left( \widehat{\mathbf{H}}_b^{\mathrm{H}} \widehat{\mathbf{H}}_b \right)^{-1},
	\label{eqn:ZF}
\end{equation}
where the diagonal matrix $\mathbf{D}_b^{\mathrm{ZF}}$ is chosen to meet the power constraint at each BS with equal UE power allocation, i.e., $\Vert \mathbf{w}_{k,b}^{\mathrm{ZF}}\Vert^2 = P_{\mathrm{B}}/N_{\mathrm{K},b}$ $\forall k, b$.

%\subsection{DL SINR and Throughput Formulation}

The DL SINR on a given time-frequency resource for UE $k$ can be calculated as
\begin{equation}
\gamma_{k,b}^{\mathrm{ZF}} =
\frac{ \left| \mathbf{h}_{k,b}^{\mathrm{H}} \mathbf{w}_{k,b}^{\mathrm{ZF}} \right|^{2}}
{\sum\limits_{j \in \mathcal{B}} \sum\limits_{k \in \mathcal{K}_j} \vert \mathbf{h}_{k,j}^{\mathrm{H}} \mathbf{w}_{k,j}^{\mathrm{ZF}} \vert^{2} \!+\! \sigma^{2}}.
\label{eq:SINR_ZF}
\end{equation}
Then, the sum DL throughput at BS $b$ can be computed as
\begin{equation} \label{Equation:spectralEfficiency}
\Gamma_{b} = \underbrace{\Big( 1 - \dfrac{\tau}{T} \Big)}_{\substack{\text{Overhead} \\ \text{loss}}} \, \cdot \!\!\!\! \underbrace{\sum_{k=1}^{N_{\mathrm{K},b}}}_{\substack{\text{Multiplexing} \\ \text{gain}}} \!\!\!\! B \log_2(1 + \!\!\!\!\!\!\!\! \underbrace{\gamma_{k,b}^{\mathrm{ZF}}}_{\substack{\text{SINR affected} \\ \text{by UL SRS} \\ \text{contamination}}}\!\!\!\!\!\!\!),
\end{equation}
where $B$ denotes the system bandwidth.

\section{UL SRS Coordination} \label{PilotReuseScheme}

\begin{figure*}[htp]
\centering
	\includegraphics[width=2.0\columnwidth]{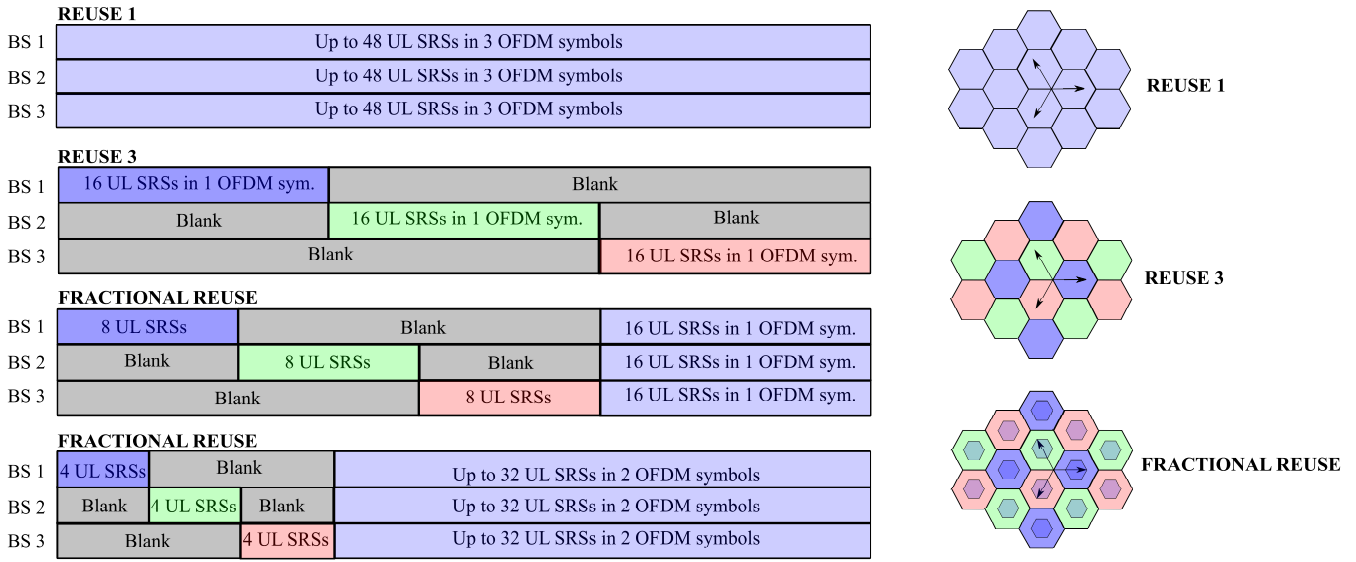}
	\caption{Example of different UL SRS allocation and reuse schemes with $\tau=3$ OFDM symbols dedicated to channel training. For each OFDM symbol, a maximum of 16 orthogonal UL SRS sequences can be allocated. The light grey area corresponds to the blanking region of the three co-located BSs, i.e., sectors, implemented for pilot interference mitigation. For FR schemes, the split between protected and shared resources is illustrated.}
	\centering
	\label{fig:ReuseSchemes}
\end{figure*}

In this section, we first recall two fixed pilot reuse schemes, namely Reuse~1 and Reuse~3. We then introduce fractional reuse, describing both the cell-centric (FR-CC) and the proposed neighbour-aware (FR-NA) approaches, along with the criteria they respectively employ to associate UEs to the pool of protected and shared UL SRS resources. Fig. \ref{fig:ReuseSchemes} illustrates all above approaches for the case when UL SRSs use $\tau=3$ OFDM symbols, thus leading to a maximum of $N_{\mathrm{P},\tau = 3}=48$ orthogonal UL SRSs. More in general, the number $N_{\mathrm{K},b}$ of scheduled UEs at the BS $b$ can be expressed as
\begin{equation} \label{eq:num_orth_sequences}
      N_{\mathrm{K},b} \leq \left\lfloor \frac{N_{\mathrm{P}, \tau}}{3 \beta_{\mathrm{PR}} + \beta_{\mathrm{SH}}} \right\rfloor.
\end{equation}
In (\ref{eq:num_orth_sequences}), $\beta_{\mathrm{PR}} \in [0,1]$ denotes the fraction of protected UL SRS sequences. These protected sequences are orthogonal among UEs served by different BSs -- i.e., in different sectors -- of the same deployment site. The remaining fraction of sequences is denoted $\beta_{\mathrm{SH}}=1- \beta_{\mathrm{PR}}$, and contains those sequences that can be reused across the three sectors of the same deployment site. We note that both Reuse~1 and Reuse~3 can be regarded as special cases of (\ref{eq:num_orth_sequences}) with $\beta_{\mathrm{PR}}=0$ and $\beta_{\mathrm{PR}}=1$, respectively.

\subsection{Reuse 1 ($\beta_{\mathrm{PR}}=0$)}  %($\beta_f=0$, $\beta=1$) 

With Reuse~1, all UEs scheduled by all BSs share the entire set of UL SRSs. 
UL SRS orthogonality is only guaranteed among UEs associated with the same BS but not among UEs connected to different BSs. 
For each OFDM symbol reserved for channel training (dark blue area in Fig. \ref{fig:FrameStructure}), $N_{\mathrm{P},\tau = 1}=16$ UL SRSs can be accommodated, 
and accordingly, $N_{\mathrm{K},b} \leq 16, \forall b \in \left\{1, \ldots, N_{\mathrm{B}} \right\}$ UEs can be multiplexed during the DL data phase by each massive MIMO BS. This scheme allows to multiplex the largest number of UEs, as all BSs can allocate all UL SRS sequences, but it yields the most severe pilot contamination. It is important to note that increasing the number of OFDM symbols dedicated to the transmission of UL SRSs, while keeping fixed the number of scheduled UEs (i.e., $N_{\mathrm{K},b}<N_{\mathrm{P}}$), allows to reduce the probability of reusing the same UL SRS across the network, thus mitigating pilot contamination.    
	
\subsection{Reuse 3 ($\beta_{\mathrm{PR}}=1$)} %($\beta_f=0$, $\beta=3$)

With Reuse~3, three co-located BS sectors within the same cell site use three different and orthogonal sets of UL SRSs. Thus, UL SRS orthogonality is preserved across the entire cell site. With respect to UL SRS Reuse 1, this scheme thus leads to lower pilot contamination. However, according to (\ref{eq:num_orth_sequences}), this effect comes at the expense of training only one third of the UEs given a fixed $\tau$,
%when considering a fixed number of UL SRSs per BS, 
or alternatively, using three times as many OFDM symbols $\tau$ when considering a fixed number of scheduled UEs per BS.
Therefore, a trade-off exists between UL SRS Reuse 1 and Reuse 3,
i.e., pilot contamination versus associated training overhead. 
It is important to note that pilot contamination can still be generated in Reuse 3 by UEs located in BSs lying in different sites that reuse the same resources. 
	
\subsection{Fractional Reuse ($0 < \beta_{\mathrm{PR}} < 1$)} \label{PartialBlanking} %($\beta_f=1/3$, $\beta=2$)

Fractional UL SRS reuse generalized fixed reuse, providing a further degree of flexibility between Reuse~1 and Reuse~3. FR applies the Reuse~3 concept for a specific fraction of UEs $\beta_{\mathrm{PR}}$ %and associated pilot sequences 
and the Reuse~1 concept for the remaining fraction $\beta_{\mathrm{SH}}$. 
For a fixed number of OFDM symbols dedicated to UL SRSs, this scheme is capable of multiplexing a larger number of sequences than Reuse~3 (but still fewer than Reuse~1) while providing less pilot contamination than Reuse~1 (but still more severe than Reuse~3). Evaluating the impact of varying $\beta_{\mathrm{PR}}$, which controls the split between protected and shared UL SRS resources, and of the way protected and shared UL SRSs are allocated to UEs can guide mobile operators on how to trade pilot contamination for overhead in their quest for higher overall capacity. We will carry out such evaluation in Sec.\ \ref{PerformanceResults}.

In what follows, we detail UL SRS allocation in two FR schemes: FR-CC and the proposed FR-NA. .
\begin{itemize}

\item {\bf Cell-centric fractional reuse (FR-CC):} 
With this approach, proposed in \cite{7247312_FractionalPilotReuseDebbah,7127059_Hanzo_SoftPilotReuse}, each BS helps its UEs positioned at the cell edge, whose UL SRSs reach the BS with low power, and who can potentially suffer more from neighbouring UEs contamination. In practice, this strategy can be implemented by ranking all scheduled UEs according to increasing values of the power $p_{i,b}$ that BS $b$ receives from each UE $i$. Such power value can be fed back to the BS via measurement reports. The protected UL SRS resources are then assigned to the top fraction $\beta_{\mathrm{PR}}$ of UEs in such ranking, whereas the shared UL SRS resources are assigned to the remaining bottom $\beta_{\mathrm{SH}}$ fraction of UEs.

\item {\bf Neighbour-aware fractional reuse (FR-NA):} 
With this strategy instead, we propose to abandon the above cell-centric approach, and to take a broader and selfless perspective by accounting for the contamination effect produced by each UE to the UL SRSs of UEs in neighbouring cells. In other words, we propose to assign the protected UL SRSs to the UEs that generate the most interference to neighbouring BSs. This strategy is implemented by ranking all scheduled UEs according to increasing values of the maximum power $\max_{j \in \mathcal{B} \backslash b} \{p_{i,j}\}$ that UE $i$ receives from all other BSs. Such power value can be reported by each UE $i$ to BS $b$ and, owing to channel reciprocity, it provides information on the interference caused by UE $i$ to all other BSs. In FR-NA, the protected UL SRS resources are then assigned to the top fraction $\beta_{\mathrm{PR}}$ of UEs in such ranking, whereas the shared UL SRS resources are assigned to the remaining bottom $\beta_{\mathrm{SH}}$ fraction of UEs.
\end{itemize}

\section{Performance Evaluation} \label{PerformanceResults}

In this section, we evaluate the performance of the schemes described in Sec.\ \ref{PilotReuseScheme} in a practical cellular deployment scenario, whose parameters are summarized in Table \ref{table:parameters}. 

\begin{table}
\centering
\caption{System-level simulation parameters}
\vspace{-0.2cm}
\label{table:parameters}
\begin{tabulary}{\columnwidth}{ |p{2.8cm} | p{5cm} | }
\hline
    \textbf{Parameter} 			& \textbf{Description} \\ \hline
    Cellular layout				& Hexagonal with wrap-around, 19 sites, 3 sectors each, 1 cellular BS per sector \\ \hline
    BS inter-site distance 		& 500~m \cite{3GPP36814} \\ \hline
	UEs distribution			& Uniform \\ \hline   %Random (P.P.P.) deployment
	Path loss and LOS prob.	& 3GPP UMa \cite{3GPP36814}  \\ \hline
	Shadowing 						& Log-normal \cite{3GPP36814} \\ \hline
	Fast fading  					& Ricean, distance-dependent K factor \cite{3GPP25996} \\ \hline
	Thermal noise 					& -174 dBm/Hz spectral density\\ \hline
	Carrier frequency 			& 2 GHz  \cite{3GPP36814} \\ \hline
	System bandwidth 			& 20 MHz with 100 resource blocks \cite{3GPP36814} \\ \hline
	Duplexing mode				& TDD \\ \hline
	BS precoder 				& Zero forcing \cite{1261332_ZF} \\ \hline
	BS antennas					& Downtilt: $12^{\circ}$, height: $25$~m  \cite{3GPP36814} \\ \hline
	BS antenna array 			& Uniform linear, element spacing: $d = 0.5\lambda$ \\ \hline
	BS antenna pattern 			& Antennas with half-power beamwidth of $65^{\circ}$ and gain of 8 dBi\ \cite{3GPP36873} \\ \hline
	BS tx power 				& 49 dBm \cite{3GPP36814} \\ \hline
	UE antennas 				& Single omnidirectional antenna \cite{3GPP36814} \\ \hline
	UE tx power 				& $23$ dBm \cite{3GPP36814} \\ \hline
	UE noise figure 			& 9 dB \cite{3GPP36814} \\ \hline
	Traffic model				& Full buffer \\ \hline
\end{tabulary}
\end{table}

\subsection{Contamination vs. Overhead Trade-off}

\begin{figure}[t]
	\includegraphics[width=\columnwidth]{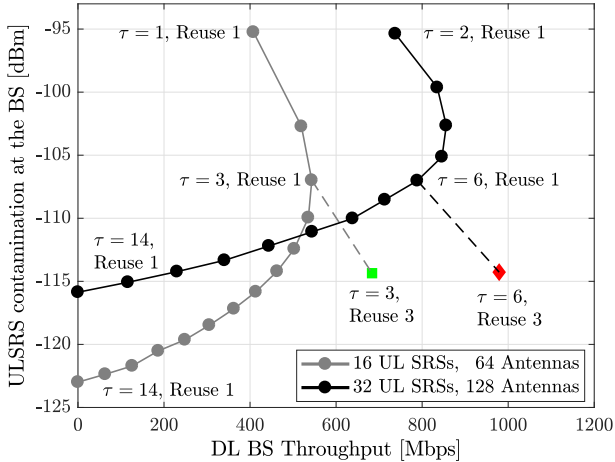}
	\caption{Trade-off between UL SRS contamination mitigation and overhead in terms of the $50^{\mathrm{th}}$- percentile of the DL BS throughput.}
	\centering
	\label{fig:64vs128Ant_SchemeAC_BS_T_withCSIOH_CDF_Median}
\end{figure}

In Fig. \ref{fig:64vs128Ant_SchemeAC_BS_T_withCSIOH_CDF_Median}, 
we evaluate the trade-off between pilot contamination %\textit{decontamination} 
and associated overhead when progressively increasing the portion $\tau$ of OFDM symbols reserved to UL SRSs, while keeping fixed the number $N_{{\mathrm{K}},b}$ of UEs scheduled. 
In particular, we show the UL SRS contamination perceived at the BS as a function of the $50^{\mathrm{th}}$-percentile values of the BS throughput when using $N_{{\mathrm{A}}} = 64$ (resp. 128) BS antennas with $N_{{\mathrm{K}},b} = 16$ (resp. 32) scheduled UEs. The value of this contamination is the measure of the interference over the received pilot sequences in (\ref{eq:pilot}) generated by UEs located in different cells and reusing the same sequences that, during the channel estimation procedure in (\ref{eq:channel_estimation}), results in the inability of separating their different channel directions. The results in this figure are obtained with Reuse~1 and with $\tau$ varying from $\tau =1$ to $\tau = 14$ (corresponding to the maximum number $T$ of OFDM symbols in the subframe). Intuitively, increasing $\tau$ translates to a reduction in the probability that the same pilot sequence is reused in neighbouring cells.
The idea behind this figure is to guide the system designer by showing the trade-off between overhead and contamination effects.

Fig.\ \ref{fig:64vs128Ant_SchemeAC_BS_T_withCSIOH_CDF_Median} shows that increasing the value of $\tau$ is initially beneficial to decrease contamination and, as a result, to augment the BS throughput. This happens mainly because the reduction of pilot contamination is predominant over the increased overhead. However, further increasing $\tau$ beyond a certain point does not increase the BS throughput, as the larger training overhead outweighs the benefits of mitigating pilot contamination. 
Fig.~\ref{fig:64vs128Ant_SchemeAC_BS_T_withCSIOH_CDF_Median} also illustrates the performance of Reuse~3, which enforces full orthogonalisation among the three co-located BSs lying in the same site. This can be achieved with $\tau=3$ for the 16 UEs case and with $\tau=6$ for the 32 UEs case.
The gain exhibited by Reuse~3 over Reuse~1 is in median around $75\%$ and $33\%$ for $N_{{\mathrm{K}},b}=16$ and $N_{{\mathrm{K}},b}=32$, respectively. Although not explicitly shown for brevity in this paper, these gains increase respectively to around $130\%$ and $70\%$ at the cell border (i.e. $5^{\mathrm{th}}$-percentile).

In practical massive MIMO deployments, systems with a larger number of simultaneously scheduled UEs represent a more challenging scenario for UL SRS coordination. In the remainder of this section, we therefore focus on the configuration with 32 UEs and 128 antennas.

\subsection{Fixed Reuse vs. Fractional Reuse}

\begin{figure}[t]
	\includegraphics[width=\columnwidth]{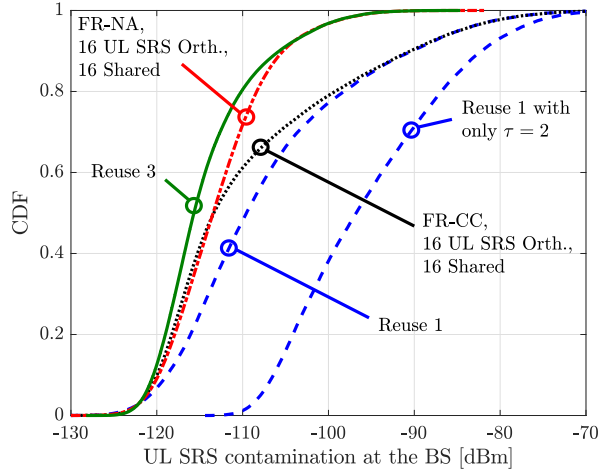}
	\caption{Distribution of the interference measured at the BS during UL SRS reception. Comparison among state-of-art reuse schemes and our proposed FR-NA method when the number of scheduled UEs is $N_{\mathrm{K},b}=32$.}
	\centering
	\label{fig:128Ant_SchemeAC_BSInterference_CDF}
\end{figure}

Fig. \ref{fig:128Ant_SchemeAC_BSInterference_CDF} represents the cumulative distribution function (CDF) of the pilot interference (i.e., contamination) measured at the BSs during the reception of UL SRSs. In this figure, we fix $\tau=6$, $N_{{\mathrm{K}},b}=32$, and $N_{\mathrm{A}} = 128$,
which results in 32 UL SRSs used by each BS. Moreover, we consider and compare the following reuse schemes: Reuse~1, Reuse~3, Cell-centric Fractional Reuse (FR-CC), and
the proposed Neighbour-Aware Fractional Reuse (FR-NA). For the cases of FR-CC and FR-NA, 
we allocate the SRSs of 16 UEs in protected time-frequency resources, whereas the remaining 16 UEs share the same pool of resources. For comparison purposes, we also illustrate the contamination incurred by Reuse~1 with $\tau=2$ OFDM symbols for CSI acquisition, which entails a $100\%$ collision probability among the UL SRSs allocated to the different BSs. 

As expected, we can see how FR-CC and FR-NA lie in between the performance of Reuse~1 and Reuse~3 in terms of contamination mitigation. Moreover, the two curves almost overlap for the lowest values of pilot interference. This area is associated with UEs that benefit of the protected resources. However, the curves diverge for higher values of pilot interference. In this area, the UEs are allocated over resources where pilot orthogonality cannot be guaranteed. The different values obtained by the curves of FR-CC and FR-NA in this region are a direct consequence of the UL SRS strategy adopted to select the UEs to be assigned protected resources. It can be seen that our proposed FR-NA scheme outperforms the FR-CC approach. Having a serving cell perspective only, as in FR-CC, turns out not to be the best approach. These results clearly indicate that one should allocate the protected UL SRSs to the UEs that can potentially create stronger interference (as in FR-NA) rather than to the ones located at the cell border (as in FR-CC).

\begin{figure}[t]
	\includegraphics[width=\columnwidth]{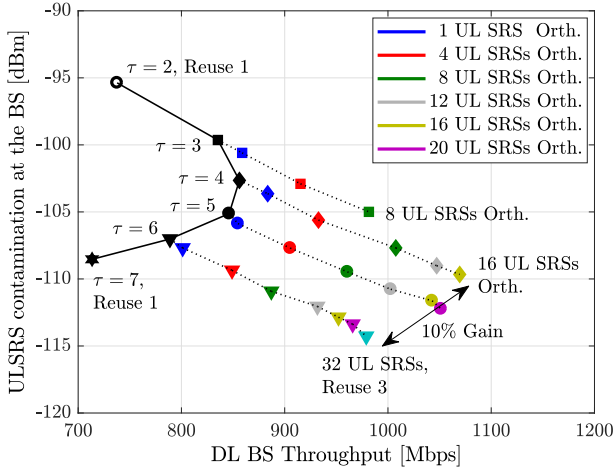}
	\caption{DL BS throughput as a function of the UL SRS contamination. Effects produced by progressive UL SRS orthogonalisation via FR-NA when $N_{\mathrm{K},b}=32$ and $N_\mathrm{A}=128$. Results measured at the $50^{\mathrm{th}}$-percentile.}
	\centering
	\label{fig:Paper5a_128Ant_Scheme_AC_BS_T_withCSIOH_Metric2_CDFMedianR1}
\end{figure}

In Fig. \ref{fig:Paper5a_128Ant_Scheme_AC_BS_T_withCSIOH_Metric2_CDFMedianR1}, 
we analyse in more detail the effects produced by our proposed FR-NA scheme when $N_{\mathrm{K},b} = 32$ and $N_{\mathrm{A}} = 128$. As in Fig. \ref{fig:64vs128Ant_SchemeAC_BS_T_withCSIOH_CDF_Median}, here results are shown for various values of $\tau=\{3,4,5,6\}$. The idea behind this figure is to guide the system designer by showing the impact of $\tau$ on the BS sum throughput in \eqref{Equation:spectralEfficiency}. It can be observed here that progressively increasing the orthogonality of UL SRS sequences from Reuse~1 (up to Reuse~3 where possible) using FR-NA, by configuring the fraction of protected resources, is effective to improve the overall performance. 
In Fig.\ \ref{fig:Paper5a_128Ant_Scheme_AC_BS_T_withCSIOH_Metric2_CDFMedianR1}, different marker colors correspond to different numbers of orthogonal UL SRS sequences, increasing from left to right. 
In particular, we can highlight a throughput gain of around $45\%$ achieved by the configuration ($\tau=4$, FR-NA) with 16 UL SRS allocated over the protected resources with respect to the configuration ($\tau=2$, Reuse~1). Moreover, the throughput gain of ($\tau=4$, FR-NA) with respect to ($\tau=6$, Reuse 3) is around $10\%$.

Although these gains may not seem remarkable, it is important to highlight that the ongoing 5G standardization may not grant $\tau=6$ OFDM symbols for the UL SRS region, 
thus compromising the ability of Reuse 3 to multiplex a high number of UEs and to fully exploit the benefits of massive MIMO BSs. 
When $\tau$ is constrained to 4 or less OFDM symbols,
the advantage of FR-NA becomes more significant, especially when it is required to multiplex a large number of UEs with the best possible channel estimation capabilities.

\subsection{Spatial Multiplexing Capability}

\begin{figure}[t]
	\includegraphics[width=\columnwidth]{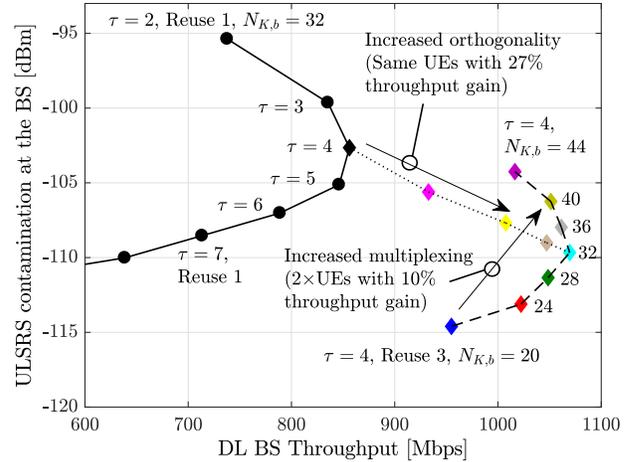}
	\caption{Interference on UL SRS sequence as a function of DL BS throughput measured as the $50^{\mathrm{th}}$-percentile. The curves show the effect of progressively increasing the number of UEs while decreasing their orthogonality.}
	\label{fig:Paper6a_128Ant_Scheme_ACD_BS_T_withCSIOH_Metric2_CDFMedian}
	\centering	
\end{figure}

In Fig.~\ref{fig:Paper6a_128Ant_Scheme_ACD_BS_T_withCSIOH_Metric2_CDFMedian}, 
we analyse the effects of spatial multiplexing while $\tau=4$, $N_{\mathrm{P},\tau=4}=64$, and $N_{\mathrm{K},b}$ varies. 
The figure is divided in two main subsets of results highlighted by arrows: 
\begin{enumerate}
\item We start from Reuse~3, which according to (\ref{eq:num_orth_sequences}) allows to accommodate $N_{\mathrm{P}}=N_{\mathrm{K},b}=20$ fully orthogonal UL SRS sequences in three co-located BSs,
and we progressively increase $N_{\mathrm{K},b}$ up to 44 %the number of UEs (i.e. UL SRSs) 
using the proposed FR-NA approach. To be able to multiplex 44 UEs, we select 10 UEs to be allocated protected resources, while we assign shared resources to the remaining 34 UEs. 

\item We keep fixed $N_{\mathrm{K},b}=32$ %the number of UL SRSs, 
and we progressively increase the orthogonality of the UL SRS sequences, thus reducing the contamination effect, by increasing the fraction of protected resources $\beta_{PR}$ to accommodate the transmission of UL SRSs for $\{4,8,16\}$ UEs in the protected resources 
(similarly to what shown in Fig.~\ref{fig:Paper5a_128Ant_Scheme_AC_BS_T_withCSIOH_Metric2_CDFMedianR1} and introduced here as a term of comparison).
\end{enumerate}

The rationale behind Fig.~\ref{fig:Paper6a_128Ant_Scheme_ACD_BS_T_withCSIOH_Metric2_CDFMedian} is to guide the system designer in quantifying the trade-off between multiplexing gain and increased contamination. The results of this figure illustrate that, given a fixed number of OFDM symbols reserved for UL SRS transmission (in this case $\tau=4$), it is possible to opt for keeping fixed the number of scheduled UEs (in this case $N_{\mathrm{K},b}=32$) and increase the orthogonality to the maximum level, providing a $26\%$ throughput gain with respect to Reuse~1. Moreover, Fig. \ref{fig:Paper6a_128Ant_Scheme_ACD_BS_T_withCSIOH_Metric2_CDFMedian} shows that the proposed FR-NA makes it possible to double the number of spatially multiplexed UEs with respect to Reuse~3 (from 20 to 40), while still providing a $10\%$ throughput gain thanks to both $i)$ the selection of the appropriate number of time-frequency resources allocated to the protected UL SRSs, and $ii)$ the proposed FR-NA UE selection for pilot sequence assignment. 

\section{Conclusion} \label{Conclusions}

We compared the virtues and shortcomings of various UL SRS allocation strategies with the aim of providing useful takeaways for the practical design of massive MIMO systems. While the detrimental effect of pilot contamination can be reduced through conservative pilot reuse schemes, their benefits vanishes when the correspondent overhead is considered. As a trade-off between conservative and aggressive fixed reuse approaches, we proposed a novel fractional reuse strategy based on selflessly relieving neighbouring BSs of severe pilot contamination. The results of extensive simulation campaigns prompted us to draw the following conclusions: 
\emph{i)} If the number of OFDM symbols $\tau$ dedicated to UL SRS is large, then Reuse~3 is the solution of choice; 
\emph{ii)} If only a relatively small $\tau$ is allowed instead -- likely to occur in practice --, then Reuse~3 becomes infeasible, and fractional UL SRS reuse can be seen as the viable alternative;
\emph{iii)} Allocating protected UL SRS resources to UEs that create the most contamination to neighbouring cells -- as proposed -- outperforms a cell-centric approach that simply protects cell-edge UEs.

\ifCLASSOPTIONcaptionsoff
  \newpage
\fi
\bibliographystyle{IEEEtran}
\bibliography{bibfile}
\end{document}